\documentclass[journal]{IEEEtran}
\usepackage{hyperref}
\usepackage[hyphenbreaks]{breakurl}
\usepackage{xurl}
\usepackage[nocompress]{cite}
\usepackage{amsmath,amssymb,amsfonts}
\usepackage{graphicx}
\usepackage{textcomp}
\usepackage{microtype}
\usepackage{subcaption}
\usepackage{booktabs}
\usepackage{multirow}
\newcommand{\method}{\texttt{PromptGraph}}
\newcommand{\mask}{\texttt{[MASK]}}

\title{\method: Graph-Guided Prompt Sanitization for Balancing Privacy and Utility in LLM Inference}
\author{Chen Gu, Hui Wan, Donghui Hu, Hui Wang, and Zhuoer Gu%
\thanks{Chen Gu, Hui Wan, Donghui Hu, and Hui Wang are with the School of Computer Science and Information Engineering, Hefei University of Technology, Hefei, China (e-mail: guchen@hfut.edu.cn; wanhui@mail.hfut.edu.cn; hudh@hfut.edu.cn; wanghui@hfut.edu.cn).}%
\thanks{Zhuoer Gu is with International College Beijing, China Agricultural University, China (e-mail: guzhuoer@cau.edu.cn).}}
\hypersetup{
    hidelinks,
    pdftitle={PromptGraph: Graph-Guided Prompt Sanitization for Balancing Privacy and Utility in LLM Inference},
    pdfauthor={Chen Gu, Hui Wan, Donghui Hu, Hui Wang, Zhuoer Gu}
}

\begin{document}

\maketitle

\begin{abstract}
Large Language Model (LLM) services introduce a fundamental privacy challenge. Sensitive information may be inferred not only from explicit identifiers, such as names or phone numbers, but also from contextual associations among otherwise innocuous spans. Existing sanitizers typically assign privacy or utility signals to individual spans without explicitly modeling pairwise relationships among them. In this paper, we propose \method{}, a graph-guided prompt-sanitization approach for privacy-preserving LLM inference. \method{} estimates privacy leakage at the span level and utility-relevant contextual dependencies between pairs of spans. It represents each prompt as an attributed graph, in which nodes carry span-level privacy scores and edges encode contextual dependencies needed to preserve utility. The sanitization objective selects a protected span set that maximizes privacy gain while penalizing the loss of contextual dependencies. This formulation explicitly balances privacy and utility when contextual evidence is hidden. Protected spans are sanitized locally, and returned placeholders are restored only after passing local consistency checks. We conduct extensive experiments showing that \method{} achieves a more favorable balance between privacy and utility than prompt-privacy baselines.
\end{abstract}

\begin{IEEEkeywords}
Prompt sanitization, privacy-preserving inference, large language models, graph-guided selection.
\end{IEEEkeywords}

\section{Introduction}
Cloud Large Language Models (LLMs) are widely deployed through a prompt--response workflow, in which users submit prompts to remote servers and receive generated outputs \cite{ouyang2022training}. Although this paradigm makes foundation models broadly accessible, it requires prompt content to be disclosed during each remote inference request \cite{bender2021dangers}. Such prompts may contain sensitive information, including personal attributes and account credentials. Consequently, prompts transmitted to remote LLM servers may expose users’ private information.

Prior work has explored several strategies for reducing privacy leakage in prompts. Differentially private text sanitization \cite{feyisetan2020privacy,chen2022customized} perturbs tokens or representations to limit disclosure. However, these classical perturbation methods are not primarily designed for prompt workflows that require local restoration after generation. Prompt sanitization methods instead protect the input before cloud inference. HaS \cite{chen2023has} hides private entities locally and restores anonymized responses after generation. InferDPT \cite{tong2023inferdpt} studies privacy-preserving inference for black-box LLMs through perturbation and extraction. DP-OPT \cite{hong2023dpopt} investigates a related but distinct prompt optimization setting under differential privacy. Closer to our setting, ProSan \cite{shen2024fire} balances privacy leakage risk with word importance, while ALSA \cite{ma2025alsa} considers privacy leakage risk, contextual information importance, and task relevance when assigning anonymization actions. These studies suggest that lightweight local sanitization can be integrated into online LLM workflows.

\begin{figure}[t]
\centering
\includegraphics[width=0.95\columnwidth]{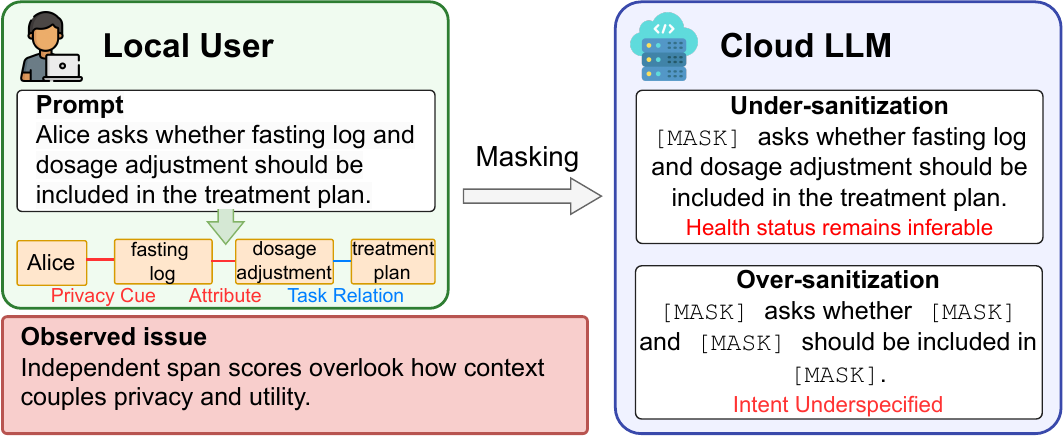}
\caption{Context-coupled privacy risks of independent span sanitization. Under-sanitization may mask an explicit value while leaving contextual cues that support sensitive-attribute inference, whereas over-sanitization may remove task-relevant relations and leave the prompt underspecified for downstream LLM inference.}
\label{fig:context-coupled}
\end{figure}

Despite this progress, existing prompt sanitization methods still do not explicitly model how privacy leakage and task utility are coupled through contextual relations. As illustrated in Figure~\ref{fig:context-coupled}, independent span decisions may hide an explicit sensitive value while leaving nearby cues that support recovery or attribute inference. They may also remove contextual relations that downstream inference needs. Existing methods such as ProSan \cite{shen2024fire} and ALSA \cite{ma2025alsa} mainly attach importance or relevance signals to individual words or spans, rather than to the relations among them.
In ALSA, for example, contextual information importance is summarized as a word-level feature and combined with privacy and task relevance in a clustering-based action rule. 
Such designs do not identify which visible spans provide evidence for recovering a sensitive or hidden span, nor do they score the utility loss caused by removing a particular relation. The key problem is therefore \textbf{\emph{how to select prompt content for protection while reducing contextual privacy risk and preserving relations needed for downstream inference}}.

In this paper, we propose \method{}, a graph-guided prompt sanitization approach that hides evidence of private information while preserving contextual relations. It first estimates span privacy from local patterns and counterfactual masking, so contextual privacy cues are handled as node-level privacy evidence. It then estimates contextual dependency for utility preservation by checking how much one span supports local reconstruction of another. This edge-level dependency goes beyond span relevance by making utility-relevant pairwise relations explicit in the selection objective. The scored spans and retained pairwise relations are organized as a prompt graph for selection. The approach selects protected spans through a principled trade-off between privacy gain and contextual utility loss. Protected spans are locally sanitized, and returned placeholders are restored only when they pass local consistency checks. This graph-guided selection protects high-risk spans while avoiding unnecessary removal of contextual relations needed for downstream tasks. The \textbf{novelty} of \method{} lies in reframing prompt sanitization from isolated span scoring as relation-aware graph selection, enabling a more explicit balance between privacy protection and task utility.

Our contributions are summarized as follows.
\begin{itemize}
    \item We propose \method{}, a graph-guided prompt sanitization approach that protects sensitive prompt content while preserving contextual relations needed for downstream inference.
    \item The approach first segments a prompt into textual spans, estimates span privacy risk and contextual dependency for utility preservation, and organizes the scored spans and relations as an attributed prompt graph. It then selects protected spans with an edge-aware objective and restores returned placeholders only after local consistency checks.
    \item We conduct extensive experiments on three task datasets, three mainstream LLMs, and two server-side attack evaluations, demonstrating a more favorable balance between privacy and utility than baselines.
\end{itemize}

\section{Related Work}

\subsection{Privacy-Preserving Prompt Solutions}
Privacy-preserving prompt solutions aim to protect sensitive information during cloud LLM inference while preserving the utility of sanitized prompts for downstream tasks. 
Practical approaches typically anonymize prompts locally and restore protected content after generation. HaS \cite{chen2023has} replaces sensitive entities before inference and restores them in the returned response. Another line of work reduces disclosure through calibrated perturbations, often under differential privacy guarantees. Early text sanitization methods \cite{feyisetan2020privacy,chen2022customized} perturb tokens or representations to limit information leakage. InferDPT \cite{tong2023inferdpt} investigates privacy-preserving inference for black-box LLMs through perturbation and extraction, while DP-OPT \cite{hong2023dpopt} applies differential privacy to prompt optimization rather than direct prompt sanitization. More recent methods aim to balance privacy and utility. ProSan \cite{shen2024fire} balances privacy leakage risk against word importance, and ALSA \cite{ma2025alsa} additionally incorporates contextual importance and task relevance when selecting anonymization actions. Nevertheless, existing approaches still make protection decisions primarily at the span level, even when contextual or task-aware signals are considered.

\subsection{Contextual Evidence and Structured Prompting}
Recent studies have shown that LLM behavior is strongly influenced by both the content and organization of contextual information. In-context learning research demonstrates that demonstration retrieval \cite{rubin2022learning}, prompt format, example selection, and example ordering can all substantially affect downstream performance \cite{zhao2021calibrate}. Studies on long-context models \cite{liu2024lost} further reveal that LLMs do not utilize contextual evidence uniformly, with performance often depending on the position of relevant information within the prompt. While these findings highlight the importance of contextual evidence for task performance, they do not address how such information should be protected or preserved during prompt sanitization.
Structured prompting provides another perspective on how contextual dependencies can be represented. Prior prompting methods elicit intermediate reasoning steps in explicit chains or organize reasoning states as graphs \cite{wei2022chain,besta2024graph}, showing that structure can support more flexible LLM reasoning. However, these structures are designed for reasoning rather than for selecting which prompt content should be hidden before cloud inference. %In contrast, \method{} uses graph structure for local prompt sanitization: nodes capture privacy leakage risk, while edges capture contextual dependencies whose removal may harm downstream inference.

\section{Preliminaries}

\subsection{Counterfactual Masking}
Counterfactual analysis \cite{wachter2018counterfactual} estimates the effect of an input component by comparing a score before and after a controlled intervention. Let $x=[u_1,u_2,\ldots,u_n]$ denote an input represented as a sequence of components, and let $T\subseteq\{1,\ldots,n\}$ be an intervention set. A masked counterfactual input is defined as:
\begin{equation}
    x_{\setminus T}
    =
    [\bar u_1,\ldots,\bar u_n],
    \quad
    \bar u_k=
    \begin{cases}
        \mask{}, & k\in T,\\
        u_k, & k\notin T.
    \end{cases}
\end{equation}

For a scoring function $H(\cdot)$, the positive counterfactual contribution of component $u_i$ can be written as:
\begin{equation}
    \Delta_i^H=[H(x)-H(x_{\setminus\{i\}})]_+ ,
\end{equation}
where $[\cdot]_+=\max(\cdot,0)$ retains only positive score reductions caused by masking. This notation quantifies the marginal effect of an intervened component on the output of an estimator, without specifying how components are selected or how the estimator is implemented.

\subsection{Threat Model}
We consider black-box LLM inference with a local sanitizer. Given an original prompt $q$, the client sends only the sanitized prompt $\tilde q$ to the cloud model, receives a sanitized response $\tilde r$, and locally restores the placeholders using a private restoration table to obtain the final response $r$.

The cloud service is honest but curious and can observe only artifacts visible to the server: the sanitized prompt $\tilde q$, the raw response $\tilde r$, and the public sanitization procedure. It does not observe the original prompt $q$, the restoration table, local detector scores, or restoration decisions on the client. Under this view, we consider two risks on the server side. A recovery attack attempts to reconstruct hidden sensitive values from visible context. An attribute inference attack attempts to infer sensitive attribute types from residual cues even when the original values remain hidden. We exclude client compromise and attacks using records specific to the user.

\subsection{Problem Formulation}
Let $q=[u_1,u_2,\ldots,u_n]$ be a prompt segmented into textual spans. A span may reveal sensitive information either directly through its surface form or indirectly through its contextual relations with other spans. Prompt sanitization aims to construct a sanitized prompt $\tilde q$ that reduces such evidence while preserving the information required for downstream inference. We formulate this problem over a prompt graph:
\begin{equation}
    \mathcal{G}(q)=(V,E,\mathbf{P},\mathbf{I}),
\end{equation}
where $V=\{v_i\}_{i=1}^n$ and each node $v_i$ corresponds to span $u_i$. The edge set $E\subseteq\{(v_i,v_j):1\le i<j\le n\}$ contains retained contextual relations, $\mathbf{P}$ stores span privacy scores $P_i$, and $\mathbf{I}$ stores utility-oriented dependency scores $I_{ij}$. A protected set $S\subseteq V$ induces a sanitized prompt $\tilde q$. Because replacing either endpoint makes the original pairwise context unavailable to the cloud, every edge incident to $S$ is affected. We therefore select $S$ to maximize privacy gain while minimizing the total weight of affected dependencies.

\section{Methodology}

\subsection{Overview}
\begin{figure*}[t]
\centering
\includegraphics[width=0.95\textwidth]{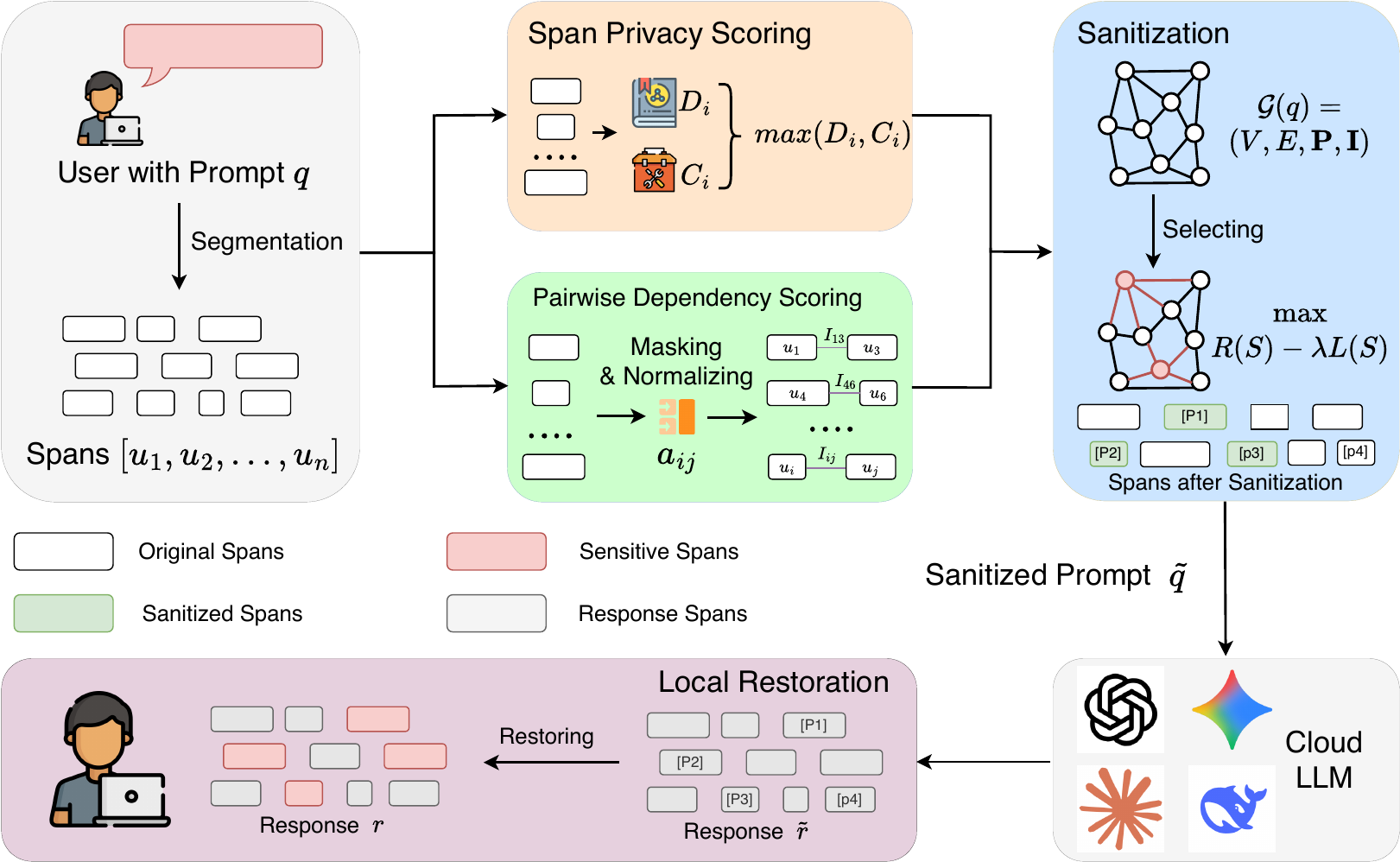}
\caption{Overview of \method{}. The local client segments the input prompt into textual spans, estimates span-level privacy risk and pairwise contextual dependency for utility preservation, and forms an attributed graph whose nodes correspond to spans and whose edges encode contextual dependencies. Graph-guided selection replaces protected spans with locally generated placeholders ([P1]–[P4]) before cloud inference. Only the sanitized prompt is sent to the cloud. The private mapping and restoration remain on the client after the response is returned.}
\label{fig:framework}
\end{figure*}

The overview of \method{} is shown in Figure \ref{fig:framework}. It operates on the client side for prompt sanitization before inference and response restoration after inference. Given an original prompt, the local sanitizer first segments it into textual spans, estimates span privacy leakage risk, and scores contextual dependency for utility preservation for retained pairs. These scored spans and pairwise relations are then organized as a prompt graph that guides protection selection. The selection algorithm then identifies protected spans and replaces each of them with an opaque placeholder carrying only a local identifier generated by the client. Only the sanitized prompt is sent to the cloud LLM. After the cloud model returns a response, the local postprocessor checks placeholder consistency before restoration. Thus the cloud service observes only the sanitized prompt, the model response, and intentionally opaque placeholder identifiers.

\subsection{Span Privacy Scoring}
Span privacy leakage risk is estimated from two complementary sources. The first is the direct evidence score $D_i$, obtained from local rules. As a surface-form score, spans that exactly match or are confidently identified by privacy detectors receive high risk values, while unmatched spans receive low or zero scores. It captures explicit sensitive patterns such as email addresses, physical addresses, and database credentials. However, direct evidence covers only surface patterns that local detectors can explicitly match and therefore cannot capture attribute evidence implied by otherwise ordinary contextual spans.

The second source relies on a local privacy estimator $G(a|x)$, which returns an exposure score for sensitive attribute $a\in\mathcal{A}$ in text $x$, where $\mathcal{A}$ denotes the set of sensitive attributes considered locally. Unlike $D_i$, which scores a single span by direct surface matches, $G$ scores the whole text and estimates attribute exposure from contextual cues. In practice, $G$ can be implemented as an ensemble of lightweight attribute classifiers and cue detectors on the client. 
This local estimator enables counterfactual assessment of contextual evidence. Beyond explicitly sensitive strings, otherwise benign spans may become privacy sensitive through their contextual associations. 
For each span $u_i$, we construct a counterfactual prompt $q_{\setminus i}$ by replacing $u_i$ with a designated placeholder. The placeholder preserves the span position while withholding its original content. We then estimate the contribution of $u_i$ by comparing attribute evidence before and after masking:
\begin{equation}
    C_i =
    \max_{a\in\mathcal{A}}
    \left[
        G(a\mid q)-G(a\mid q_{\setminus i})
    \right]_+ ,
\end{equation}
where $[\cdot]_+$ retains only positive reductions in attribute evidence. Intuitively, a span receives a high contextual score when masking it substantially reduces evidence for inferring a sensitive attribute. Consequently, spans without explicit sensitive strings may still receive high privacy scores if they make sensitive attributes easier to infer.
Because both $D_i$ and $C_i$ are on the same normalized scale, the final span privacy score combines both sources conservatively:
\begin{equation}
    P_i=\max(D_i,C_i).
\end{equation}

The maximum operator protects explicitly sensitive spans even when the contextual estimator is uncertain, while still identifying ordinary spans whose removal substantially reduces sensitive attribute evidence.

\subsection{Pairwise Dependency Scoring}

For each retained candidate pair $(u_i,u_j)$, we assign a contextual dependency score $I_{ij}$. Unlike the span privacy score $P_i$, $I_{ij}$ estimates the potential utility loss caused by removing a relation. A pair is contextually dependent when either span supports local reconstruction of the other.

To reduce computation, we score contextual dependencies only for a sparse candidate set $\mathcal{C}$ rather than for all span pairs. We construct $\mathcal{C}$ with the top $k$ nearest neighbors in the span embedding space. Each span is encoded, compared with other spans by cosine similarity, and connected to its $k$ most similar neighbors. We treat the resulting candidate pairs as undirected. This keeps the relation graph compact while retaining pairs that are likely to share useful context.

Let $\ell_i(x)$ denote a local reconstruction score that measures how strongly
the visible context $x$ supports recovery of target span $u_i$. It is a utility
signal, rather than a privacy-risk or attack score. The pairwise construction
below applies to any local scorer that produces $\ell_i(x)$. For example, a
masked language model with parameters $\theta$ provides the following
token-level instantiation \cite{salazar-etal-2020-masked}. Let
$u_i=(w_{i,1},\ldots,w_{i,m_i})$ contain $m_i$ tokens. For a prompt $x$ that
retains $u_i$ but may mask other spans, let $x_{\setminus(i,t)}$ mask only
$w_{i,t}$:
\begin{equation}
    \ell_i(x)=\frac{1}{m_i}\sum_{t=1}^{m_i}
    \log P_{\theta}\!\left(w_{i,t}\mid x_{\setminus(i,t)}\right).
\end{equation}

Length normalization makes scores comparable across spans of different
lengths. Larger $\ell_i(x)$ values indicate stronger contextual support for
recovering $u_i$.

For a retained candidate pair $(u_i,u_j)$, we measure dependency by the
reduction in one span's reconstruction score when the other is masked:
\begin{equation}
    \begin{aligned}
    a_{ij}=\frac{1}{2}\Big(
        &[\ell_i(q)-\ell_i(q_{\setminus j})]_+ \\
        &+[\ell_j(q)-\ell_j(q_{\setminus i})]_+
    \Big).
    \end{aligned}
\end{equation}

The two terms capture support in opposite directions, and the positive part
retains only utility-relevant dependencies. Contextual dependency scores are
then normalized within the retained candidate set $\mathcal{C}$:
\begin{equation}
    I_{ij}=a_{ij}/\max_{(u_p,u_q)\in \mathcal{C}}a_{pq}.
\end{equation}

If all raw scores are zero, we set $I_{ij}=0$ for every candidate pair. After normalization, larger $I_{ij}$ values indicate stronger contextual dependency. An edge is affected once either endpoint is protected and is penalized only once.

\subsection{Dependency-Aware Sanitization}

After scoring, \method{} seeks a protected set that maximizes the following objective:

\begin{equation}
    \begin{gathered}
        \max_{S\subseteq V}\;F(S),\qquad
        F(S)=R(S)-\lambda L(S),\\
        \text{where}\quad
        \left\{
        \begin{aligned}
            R(S)&=\sum_{v_i\in S}P_i,\\
            E_{\mathrm{aff}}(S)
            &=\{(v_i,v_j)\in E:
            \{v_i,v_j\}\cap S\neq\varnothing\},\\
            L(S)&=\sum_{(v_i,v_j)\in E_{\mathrm{aff}}(S)}I_{ij}.
        \end{aligned}
        \right.
    \end{gathered}
\end{equation}

Here, $\lambda$ controls the privacy--utility trade-off. Each edge is charged when its first endpoint enters $S$ and is not charged again if the other endpoint is later selected. When $\lambda=0$, the objective performs span-only sanitization. Larger $\lambda$ increasingly prioritizes contextual-dependency preservation, trading privacy gain for utility.

We approximately optimize this discrete objective with a greedy marginal heuristic. For each unprotected node, its marginal gain equals its privacy score minus the $\lambda$-weighted sum of newly affected edges to other unprotected nodes. Starting from $S=\varnothing$, \method{} repeatedly adds the node with the largest positive marginal gain and stops when none remains.

After selection, each span in $S$ is replaced with an opaque placeholder carrying a local identifier, producing the sanitized prompt $\tilde q$. Only $\tilde q$ is sent to the cloud LLM for inference, while the original spans, placeholder mapping, and private type metadata remain local.

\subsection{Local Restoration}
After the cloud LLM returns a response $\tilde r$, the local postprocessor scans it for placeholder strings. For each observed placeholder, it first checks whether its local identifier matches an entry in the private restoration table. It then applies the local consistency checks using the private metadata stored with that entry. A placeholder is restored only when both checks succeed, replacing it with the corresponding original protected span in the final response $r$. The restoration table and its metadata remain on the client throughout this process.

If the model rewrites a placeholder, invents a new identifier, or omits a placeholder, the corresponding item remains unresolved rather than being mapped to a protected span. Thus, the postprocessor does not restore a placeholder merely because it resembles a valid one. This validation confines restoration to client-side entries that pass local checks and prevents malformed or fabricated placeholders from triggering unintended restoration of sensitive content.

\section{Experiments}
\subsection{Experimental Setup}
\subsubsection{Datasets and Models}
We evaluate \method{} on three task datasets: MedQA \cite{jin2021medqa} is used for medical question answering, SAMSum \cite{gliwa2019samsum} for dialogue summarization, and CodeAlpaca-20k for instruction-style code generation \cite{chaudhary2023codealpaca}.
To evaluate response quality from LLMs, we use three mainstream downstream LLMs: Llama-3.1-8B \cite{grattafiori2024llama3}, Mistral-7B \cite{jiang2023mistral7b}, and Qwen3-8B \cite{yang2025qwen3technicalreport}.

\subsubsection{Baselines}
We compare six baselines spanning rule-based masking and representative prompt privacy methods. \textbf{RegexMasking} removes structured sensitive strings using predefined patterns. \textbf{HaS} \cite{chen2023has} performs local hide-and-seek anonymization with restoration after generation. \textbf{InferDPT} \cite{tong2023inferdpt} and \textbf{DP-OPT} \cite{hong2023dpopt} represent private inference by perturbation and differentially private prompt optimization, respectively. \textbf{ProSan} \cite{shen2024fire} balances privacy risk with word importance, while \textbf{ALSA} \cite{ma2025alsa} further incorporates contextual importance and task relevance when assigning anonymization actions.

\subsubsection{Evaluation Metrics}
Privacy is measured with Privacy Hiding Rate (PHR) \cite{shen2024fire} and two attack metrics on the server side: Recovery Attack Success Rate (Rec-ASR) \cite{carlini2021extracting} and Attribute Inference Attack Success Rate (Attr-ASR) \cite{melis2019exploiting,staab2024beyond}. Rec-ASR measures whether the attacker can recover the exact sensitive value, while Attr-ASR measures whether the attacker can infer the sensitive attribute type or meaning even without recovering the exact value.
We utilize task utility to measure whether the cloud LLM's response remains correct and useful after prompt privacy protection.
Efficiency is evaluated by the local preprocessing overhead, including Sanitization Time Cost (STC) and Peak Memory Overhead (PMO).

Please note that, because the three datasets represent distinct task types, we use task-specific utility metrics: answer accuracy for MedQA, ROUGE-L \cite{lin2004rouge} for SAMSum, and equal-weight CodeBLEU \cite{ren2020codebleu} for CodeAlpaca. We collectively refer to these measures as \textit{Task Utility}.

\subsubsection{Parameter Settings}
The main experiments set the trade-off coefficient to $\lambda=0.2$ and the embedding neighbor sparsification parameter to $k=4$. 
The direct evidence term $D_i$ is computed from the outputs of rule-based entity recognizers and custom regular expression recognizers \cite{neamatullah2008automated}. Mapped detector scores are normalized on the validation split of AI4Privacy PII-Masking-300k \cite{singh2025unmasking}, a PII masking dataset used in recent evaluations of masking models.

%Due to space constraints, we provide the experimental implementation details in the Appendix.

\subsection{Experimental Results}

\subsubsection{Privacy Evaluation}
\begin{table}[t]
\centering
\scriptsize
\setlength{\tabcolsep}{2.2pt}
\begin{tabular*}{\columnwidth}{@{\extracolsep{\fill}}llcccccc@{}}
\toprule
\multirow{2}{*}{\textbf{Model}} & \multirow{2}{*}{\textbf{Method}} & \multicolumn{4}{c}{\textbf{PHR}$\uparrow$} & \multicolumn{2}{c}{\textbf{ASR}$\downarrow$} \\
\cmidrule(lr){3-6}\cmidrule(lr){7-8}
 &  & Med & SAM & Code & Avg. & Rec & Attr \\
\midrule
\multirow{7}{*}{Llama} & RegexMasking & 0.652 & 0.662 & 0.547 & 0.620 & 0.087 & 0.700 \\
 & HaS & 0.859 & \textbf{0.884} & 0.458 & 0.734 & 0.283 & 0.705 \\
 & InferDPT & 0.759 & 0.780 & \textbf{0.898} & 0.812 & 0.078 & 0.472 \\
 & DP-OPT & 0.211 & 0.203 & 0.524 & 0.313 & 0.320 & 0.676 \\
 & ProSan & 0.653 & 0.749 & 0.547 & 0.650 & 0.094 & 0.612 \\
 & ALSA & 0.656 & 0.552 & 0.472 & 0.560 & 0.202 & 0.654 \\
 & Ours & \textbf{0.879} & 0.847 & 0.894 & \textbf{0.873} & \textbf{0.045} & \textbf{0.347} \\
\midrule
\multirow{7}{*}{Mistral} & RegexMasking & 0.652 & 0.662 & 0.547 & 0.620 & 0.089 & 0.700 \\
 & HaS & 0.869 & \textbf{0.887} & 0.391 & 0.716 & 0.298 & 0.717 \\
 & InferDPT & 0.755 & 0.784 & 0.816 & 0.785 & 0.087 & 0.477 \\
 & DP-OPT & 0.205 & 0.213 & 0.489 & 0.302 & 0.296 & 0.659 \\
 & ProSan & 0.650 & 0.750 & 0.551 & 0.650 & 0.099 & 0.616 \\
 & ALSA & 0.649 & 0.558 & 0.468 & 0.558 & 0.210 & 0.669 \\
 & Ours & \textbf{0.886} & 0.850 & \textbf{0.876} & \textbf{0.871} & \textbf{0.050} & \textbf{0.352} \\
\midrule
\multirow{7}{*}{Qwen3} & RegexMasking & 0.652 & 0.662 & 0.547 & 0.620 & 0.086 & 0.700 \\
 & HaS & 0.851 & \textbf{0.892} & 0.438 & 0.727 & 0.288 & 0.710 \\
 & InferDPT & 0.736 & 0.777 & 0.884 & 0.799 & 0.082 & 0.483 \\
 & DP-OPT & 0.216 & 0.221 & 0.505 & 0.314 & 0.330 & 0.654 \\
 & ProSan & 0.647 & 0.748 & 0.528 & 0.641 & 0.093 & 0.615 \\
 & ALSA & 0.648 & 0.565 & 0.470 & 0.561 & 0.201 & 0.645 \\
 & Ours & \textbf{0.898} & 0.848 & \textbf{0.890} & \textbf{0.879} & \textbf{0.047} & \textbf{0.345} \\
\bottomrule
\end{tabular*}
\caption{Privacy evaluation across downstream LLMs, datasets, and attacks on the server side.}
\label{tab:privacy-eval}
\end{table}

As shown in Table~\ref{tab:privacy-eval}, \method{} achieved the highest average PHR for each of the three downstream LLMs. PHR is computed over the server-visible sanitized prompt and raw response before local restoration. The identical RegexMasking values across model blocks reflect its deterministic prompt-side masking behavior in these runs. The consistently high PHR of \method{} suggests that its selected spans protect both explicit sensitive values and contextually related attribute cues. Although some baselines performed particularly well in individual settings, including HaS on SAMSum and InferDPT on CodeAlpaca with Llama-3.1-8B, their PHR performance varied more substantially across datasets. Overall, \method{} provided more stable privacy coverage across the three task formats.

Attack results showed the same trend. \method{} consistently achieved the lowest Rec-ASR and Attr-ASR, indicating reduced exposure of both exact sensitive values and attribute evidence. By comparison, RegexMasking remained susceptible to attribute inference, HaS still leaked contextual information despite its high SAMSum PHR, and InferDPT retained more attribute evidence. These results indicate that contextual span privacy scoring covered both explicit sensitive content and attribute cues, while the dependency-aware penalty helped preserve utility-relevant context during this privacy-driven selection.

\subsubsection{Utility Evaluation}

\begin{table}[t]
\centering
\scriptsize
\begin{tabular*}{\columnwidth}{@{\extracolsep{\fill}}llccc@{}}
\toprule
\multirow{2}{*}{\textbf{Model}} & \multirow{2}{*}{\textbf{Method}} & \multicolumn{3}{c}{\textbf{Task Utility}$\uparrow$} \\
\cmidrule(lr){3-5}
 &  & MedQA & SAMSum & CodeAlpaca \\
\midrule
\multirow{7}{*}{Llama} & RegexMasking & 0.171 & 0.104 & 0.510 \\
 & HaS & 0.230 & \textbf{0.157} & 0.403 \\
 & InferDPT & 0.241 & 0.042 & 0.263 \\
 & DP-OPT & 0.241 & 0.088 & 0.439 \\
 & ProSan & 0.246 & 0.081 & 0.459 \\
 & ALSA & 0.234 & 0.103 & 0.321 \\
 & Ours & \textbf{0.248} & 0.149 & \textbf{0.536} \\
\midrule
\multirow{7}{*}{Mistral} & RegexMasking & 0.332 & 0.202 & \textbf{0.489} \\
 & HaS & 0.320 & 0.211 & 0.442 \\
 & InferDPT & 0.154 & 0.045 & 0.292 \\
 & DP-OPT & 0.327 & 0.194 & 0.457 \\
 & ProSan & 0.304 & 0.168 & 0.444 \\
 & ALSA & 0.318 & 0.154 & 0.380 \\
 & Ours & \textbf{0.335} & \textbf{0.234} & 0.472 \\
\midrule
\multirow{7}{*}{Qwen3} & RegexMasking & 0.333 & 0.141 & \textbf{0.619} \\
 & HaS & 0.320 & 0.210 & 0.540 \\
 & InferDPT & 0.113 & 0.018 & 0.251 \\
 & DP-OPT & 0.331 & 0.145 & 0.570 \\
 & ProSan & 0.326 & 0.163 & 0.562 \\
 & ALSA & 0.289 & 0.170 & 0.381 \\
 & Ours & \textbf{0.336} & \textbf{0.240} & 0.592 \\
\bottomrule
\end{tabular*}
\caption{Utility evaluation across various LLM task models and datasets.}
\label{tab:utility-eval}
\end{table}

\begin{figure}[t]
\centering
\begin{subfigure}[t]{0.49\columnwidth}
\centering
\includegraphics[width=\linewidth]{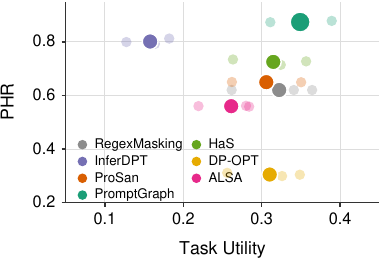}
\caption{PHR--utility trade-off}
\label{fig:privacy-utility-pareto}
\end{subfigure}
\hfill
\begin{subfigure}[t]{0.49\columnwidth}
\centering
\includegraphics[width=\linewidth]{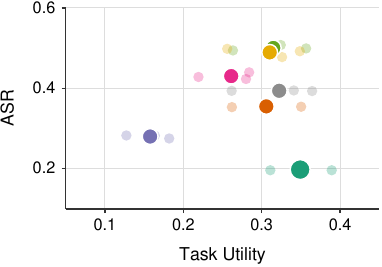}
\caption{ASR--utility trade-off}
\label{fig:utility-asr-tradeoff}
\end{subfigure}
\caption{Privacy--utility trade-offs. Faint points indicate individual downstream task models, and larger points indicate method averages. ASR is computed as the average of Rec-ASR and Attr-ASR.}
\label{fig:privacy-utility-tradeoffs}
\end{figure}

The utility evaluation tests whether the privacy gains above came at the cost of downstream task performance. As shown in Table~\ref{tab:utility-eval}, \method{} achieved the best score in six of the nine model-dataset settings and remained close to the best baseline in the other three. This stable utility is especially clear on MedQA, where \method{} was best across all models, and on SAMSum, where preserving high-dependency speaker and event relations kept summaries competitive. On CodeAlpaca, RegexMasking remained strong because many programming instructions survived simple masking, but \method{} still achieved the best score and stayed close on the other two models. These results indicate that the privacy gains of \method{} were not obtained by excessive removal of task-relevant content.
 
Based on the results in Tables~\ref{tab:privacy-eval} and~\ref{tab:utility-eval}, Figure~\ref{fig:privacy-utility-tradeoffs} summarizes the trade-off between privacy and utility. Figure~\ref{fig:privacy-utility-pareto} shows that \method{} occupies the desirable high-privacy, high-utility region, achieving the highest average PHR while maintaining the highest average task utility among all compared methods. The results in Figure~\ref{fig:utility-asr-tradeoff} show that \method{} also achieves the lowest average ASR without sacrificing utility. These results demonstrate that \method{} strengthens privacy protection while preserving task utility.

\subsubsection{Ablation Studies}
\begin{table}[!t]
\centering
\scriptsize
\setlength{\tabcolsep}{2.0pt}
\renewcommand{\arraystretch}{1.18}
\begin{tabular*}{\columnwidth}{@{\extracolsep{\fill}}llccc@{}}
\toprule
\textbf{Metric} & \textbf{Setting} & \shortstack{\textbf{Full}\\\method{}} & \shortstack{\textbf{w/o contextual}\\\textbf{term} $\boldsymbol{C_i}$} & \shortstack{\textbf{w/o pairwise}\\\textbf{weights} $\boldsymbol{I_{ij}}$} \\
\midrule
\multirow{3}{*}{PHR$\uparrow$} & MedQA & 0.879 & 0.875 & 0.940 \\
 & SAMSum & 0.847 & 0.808 & 0.853 \\
 & CodeAlpaca & 0.894 & 0.750 & 0.985 \\
 %& Avg. & 0.873 & 0.819 & 0.980 \\
\midrule
\multirow{2}{*}{ASR$\downarrow$} & Rec-ASR & 0.045 & 0.233 & 0.021 \\
 & Attr-ASR & 0.347 & 0.555 & 0.238 \\
\midrule
\multirow{3}{*}{\shortstack[l]{Task\\Utility$\uparrow$}} & MedQA & 0.248 & 0.252 & 0.117 \\
 & SAMSum & 0.149 & 0.197 & 0.057 \\
 & CodeAlpaca & 0.536 & 0.650 & 0.514 \\
\bottomrule
\end{tabular*}
\caption{Ablation of \method{}'s contextual privacy term and pairwise dependency weights using Llama-3.1-8B.}
\label{tabablation}
\end{table}

To disentangle the contributions of the two graph-aware signals, we ablate each component and observe complementary effects on privacy and utility. In Table~\ref{tabablation}, removing the contextual term $C_i$ lowered PHR and raised both attack success rates, while task utility increased. This setting still protects direct detector hits but no longer assigns privacy value to ordinary spans through counterfactual attribute evidence. The result is therefore consistent with $C_i$ reducing inferential leakage at the cost of hiding some context that supports task execution.

By contrast, eliminating the pairwise weights $I_{ij}$ strengthened measured privacy but reduced utility, with the clearest drops on MedQA and SAMSum. Because the edge penalty then vanishes from the greedy marginal gain, selection is driven only by span privacy scores and becomes more aggressive. It consequently removes additional sensitive evidence, but can also break relationships that the downstream model uses. The full method retains both signals, thereby avoiding the two one-sided behaviors: high contextual exposure without $C_i$ and excessive context removal without $I_{ij}$.

\subsubsection{Sensitivity Analysis of Hyperparameters}
\begin{figure}[!t]
\centering
\begin{subfigure}[t]{0.49\columnwidth}
\centering
\includegraphics[width=\linewidth]{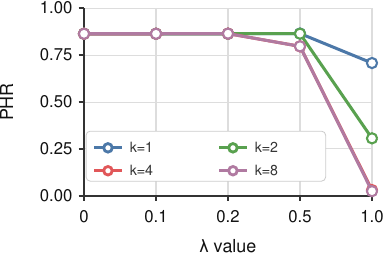}
\caption{PHR}
\end{subfigure}
\hfill
\begin{subfigure}[t]{0.49\columnwidth}
\centering
\includegraphics[width=\linewidth]{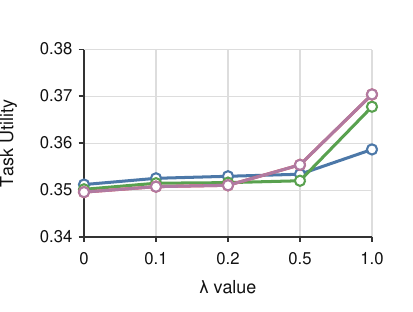}
\caption{Task Utility}
\end{subfigure}
\caption{Effects of the trade-off coefficient $\lambda$ and graph sparsity $k$ on \method{}'s privacy and task utility.}
\label{fig:hyperparameter-sensitivity}
\end{figure}

We assessed the sensitivity of \method{} to the trade-off coefficient $\lambda$ and graph sparsity $k$. Figure~\ref{fig:hyperparameter-sensitivity} reveals a broad stable region as $\lambda$ increases from 0 to 0.2, with nearly unchanged PHR and task utility across sparsity settings. The main configuration, with a coefficient of 0.2 and $k$ set to 4, lies within this plateau. Beyond this region, larger coefficients shift selection toward retaining dependency-weighted context. When$\lambda$ is 1.0, the edge-dependency penalty causes settings to select almost no content for protection. Privacy declined first for denser graphs, whereas utility improved only marginally until the largest coefficient, where privacy degradation became pronounced. Thus, the sensitivity analysis identifies a moderate trade-off coefficient as the most robust operating regime.

\subsubsection{Overhead} 
\begin{figure}[!t]
\centering
\begin{subfigure}{0.49\columnwidth}
\centering
\includegraphics[width=\linewidth]{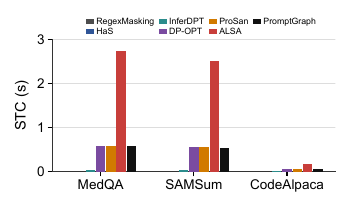}
\caption{STC Results}
\end{subfigure}
\begin{subfigure}{0.49\columnwidth}
\centering
\includegraphics[width=\linewidth]{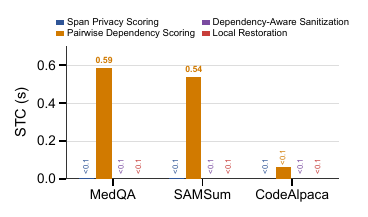}
\caption{Stage STC Breakdown}
\end{subfigure}
\centering
\begin{subfigure}{0.49\columnwidth}
\centering
\includegraphics[width=\linewidth]{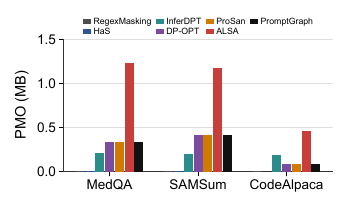}
\vspace{0.01pt}
\caption{PMO Results}
\end{subfigure}
\begin{subfigure}{0.49\columnwidth}
\centering
\includegraphics[width=\linewidth]{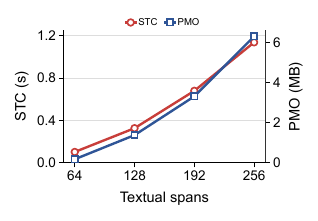}
\caption{Impact of Span Count}
\end{subfigure}
\caption{Overhead Results. PMO measures the additional memory used during sanitization, excluding resident local-model parameters.}
\label{fig:overhead}
\end{figure}

Finally, we assessed the local preprocessing overhead of \method{}. As shown in Figure~\ref{fig:overhead}(a), its sanitization time cost was comparable to that of DP-OPT and ProSan, while remaining consistently below ALSA, particularly on MedQA and SAMSum. The stage breakdown in Figure~\ref{fig:overhead}(b) identifies pairwise dependency scoring as the dominant preprocessing cost across all three datasets. Span privacy scoring, dependency-aware sanitization, and local restoration each contributed comparatively little. Figure~\ref{fig:overhead}(c) further shows that PMO, excluding resident local-model memory, remained below 0.5 MB across datasets, matching DP-OPT and ProSan and staying well below ALSA.

Moreover, Figure~\ref{fig:overhead}(d) tracks the effect of prompt length. As the number of textual spans increased from 64 to 256, both STC and PMO rose monotonically, reaching approximately 1.1 s and 6 MB at the largest setting. This trend is consistent with the growth in candidate span pairs required for dependency scoring. Overall, the overhead is concentrated in contextual pair construction rather than in the subsequent selection or restoration steps.

\section{Conclusion}
This paper introduced \method{}, a graph-guided prompt sanitizer for privacy-preserving LLM inference. \method{} selects protected spans over a graph whose nodes capture privacy risks and whose edges capture contextual dependencies needed for utility, reducing privacy exposure through contextual span scoring while preserving task-relevant relations. Across diverse datasets and task models, it improves the balance between privacy and utility by increasing PHR and reducing attack success rates while preserving competitive utility. Future work will extend this graph formulation from single-turn prompts to multi-turn interactions, where privacy evidence and useful contextual dependencies evolve across dialogue turns.

\bibliographystyle{IEEEtran}
\bibliography{references}

\end{document}